\documentclass[refree]{raa}           
\usepackage{graphicx}
\usepackage{natbib}
\usepackage{amssymb,amsmath}
\bibpunct{(}{)}{;}{a}{}{,}

\usepackage[a4paper=true,pagebackref=true]{hyperref}
\hypersetup{pdftitle = The title of my PDF, pdfauthor = My name, pdfsubject= The subject, pdfkeywords = keyword1 keyword2 keyword3} 
\hypersetup{colorlinks = true, linkcolor = green, anchorcolor = red, citecolor = blue, filecolor = red, pagecolor = red, urlcolor = red}

\begin{document}

   \title{Enhanced Remote Astronomical Archive System Based on the File-Level Unlimited Sliding-Window Technique}

 \volnopage{ {\bf 20XX} Vol.\ {\bf X} No. {\bf XX}, 000--000}
   \setcounter{page}{1}

   \author{Cong-Ming Shi\inst{1,2}, Hui Deng\inst{1}, Feng Wang\inst{1}, Ying Mei\inst{1}, Shao-Guang Guo\inst{3}, Chen Yang\inst{1}, Chen Wu\inst{4}, Shou-Lin Wei\inst{5}, Andreas Wicenec\inst{4}
   }

 \institute{ Center For Astrophysics, Guangzhou University, Guangzhou 510006, China; {\it fengwang@gzhu.edu.cn}\\
    \and
     School of Software Engineering, Anyang Normal University, Anyang 455000, China\\
	\and
    Shanghai Astronomical Observatory, Chinese Academy of Sciences, Shanghai 200030, China\\
    \and
    International Center for Radio Astronomy Research (ICRAR), The University of Western Australia, Crawley, Perth, WA, Australia\\
    \and
    Kunming University Of Science And Technology, Kunming 650500, China\\   
\vs \no
   {\small Received 2012 June 12; accepted 2012 July 27}
}

\abstract{
Data archiving is one of the most critical issues for modern astronomical observations. 
With the development of a new generation of radio telescopes, the transfer and archiving of massive remote data have become urgent problems to be solved. Herein, we present a practical and robust file-level flow-control approach, called the Unlimited Sliding-Window (USW), by referring to the classic flow-control method in TCP protocol. 
Basing on the USW and the Next Generation Archive System (NGAS) developed for the Murchison Widefield Array telescope, we further implemented an enhanced archive system (ENGAS) using ZeroMQ middleware. The ENGAS substantially improves the transfer performance and ensures the integrity of transferred files.
In the tests, the ENGAS is approximately three to twelve times faster than the NGAS and can fully utilize the bandwidth of network links. 
Thus, for archiving radio observation data, the ENGAS reduces the communication time, improves the bandwidth utilization, and solves the remote synchronous archiving of data from observatories such as Mingantu spectral radioheliograph. It also provides a better reference for the future construction of the Square Kilometer Array (SKA) Science Regional Center.
}

   \authorrunning{Cong-Ming Shi et al. }            
   \titlerunning{Enhanced Remote Astronomical Archive System}
   \maketitle
    `

%
\section{Introduction}           
\label{sec:intro}

Modern astronomical observatories or stations are generally located in sparsely populated areas with specific enviromental conditions, dictating the remote transfer of large amounts of observational data to specific data processing centers. Therefore, building a reliable system for transferring and archiving data is essential for modern astronomical telescope systems~\citep{dewdney2013ska1}. 
For example, 
the Square Kilometer Array (SKA) telescope\citep{schilizzi2004square, dewdney2009square} will be built in radio-quiet zones in two host countries (the Karoo site in South Africa and the Boolardy site in Western Australia).
During its first phase (SKA1), the SKA will generate 50$\sim$300PB of archival data per year; 
during the second phase (SKA2), the volume of newly added archival data will increase by approximately 100 times\citep{chrysostomou2018square}. 
To tackle the data deluge expected from the SKA observatory and enable the community to exploit SKA data for high impact science communication, the massive volumes of data generated by SKA will have to be transferred to SKA Regional Centres (SRCs) worldwide\citep{barbosa2020portuguese,an2019ska} via a 100 Gbit/s network link. 
The Five-hundred-meter Aperture Spherical radio Telescope (FAST)\citep{nan2011five} also plans to transfer massive filterbank data in real-time from the FAST observatory to two early science centers near Guiyang city, Guizhou via a 100 Gbit/s fiber link\citep{li2018fast}.

The Next Generation Archive System (NGAS) is a robust remote HTTP-based data archive system that has been widely deployed. It was initially developed by the European Southern Observatory (ESO)\citep{wicenec2001eso}. 
The SKA precursor facility, Murchison Widefield Array (MWA), uses the NGAS to synchronize the mirrored archive data from MWA to the data centers at the Massachusetts Institute of Technology (MIT), United States, and the Victoria University of Wellington (VUW), New Zealand\citep{wu2013optimising}. In addition, the NGAS is used by the Atacama Large Millimeter/submillimeter Array (ALMA) to synchronize massive amounts of archive data from ALMA to the data centers located in North America, Europe, and the East Asia\citep{wicenec2010alma, stoehr2014alma}.
The NGAS can deliver significantly superior file-transfer performance. Deploying multiple NGAS Subscribers and NGAS Providers is an easy and effective approach to improve the file-transfer performance. The maximum throughput that the NGAS achieved running 24 clients(Subscribers) against four servers(Providers) on a total of 28 machines was 1100 MB/s, which significantly exceeds the required rate\citep{wicenec2012mwa}. 

In developing the data archiving system for the Mingantu Spectral Radioheliograph (MUSER) and investigating the SRCs' data exchange solutions, we analyzed the source program of the NGAS and tested it experimentally. 
We demonstrated that the NGAS is an excellent candidate for future remote data archiving systems, but there is still potential for improvement and optimization.

The rest of the paper is organized as follows. In section \ref{sec:motivation}, we analyze the existing deficiencies of the NGAS. Then, we introduce our enhanced high-performance NGAS (ENGAS) in section~\ref{sec:e_ngas}, test its performance in section~\ref{sec:experiment}, present and discuss our results in section~\ref{sec:discussion}, and finally conclude our paper in section~\ref{sec:conclusion}.

\section{NGAS Framework Analysis and Transfer Performance Evaluation}\label{sec:motivation}

\subsection{NGAS Framework Analysis}
\begin{figure*}[htbp]
\centering
\includegraphics[width=0.8\textwidth]{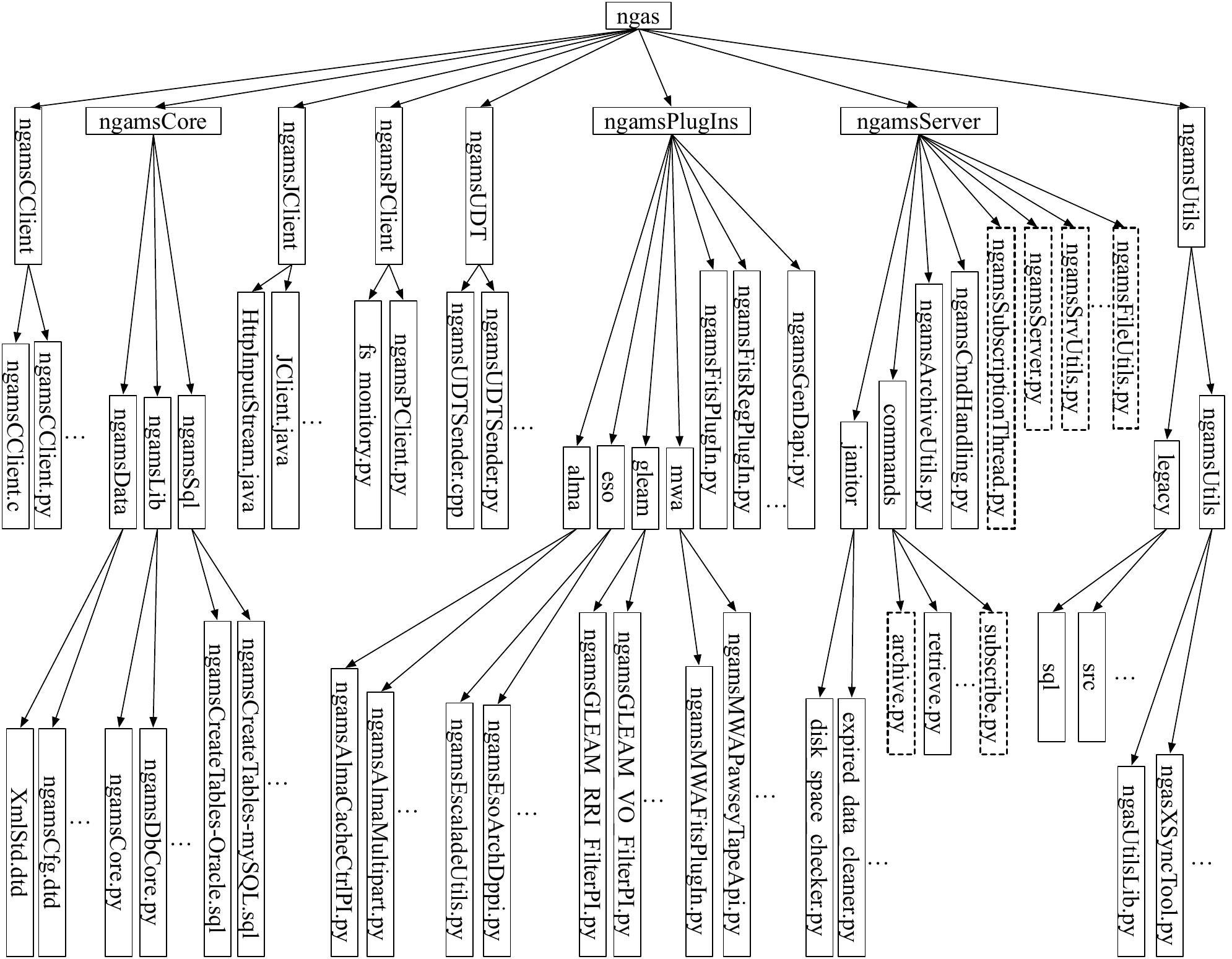}
\caption{A module diagram of NGAS}\label{ngasModuleOrganization}
\end{figure*}

The NGAS was developed in pure Python language and has a high degree of portability. Moreover, it
 is a highly modular application system.
A schematic module diagram of the NGAS is shown in Figure~\ref{ngasModuleOrganization}.
The NGAS contains eight modules: Fundamental library (ngamsLib), NGAS C-Client (ngamsCClient), NGAS Java-Client(ngamsJClient), NGAS Python-Client (ngamsPClient), Plug-In support(ngamsPlugIns), Server Module(ngamsServer), UDP-based data transfer module (ngamsUDT), and Utils module (ngamsUtils).

The NGAS Server, which is a multi-thread HTTP server based on the HTTP POST, GET, and PUT methods, is the core of the NGAS.
At least 20 application layer commands supported by the NGAS server are mainly used to archive, retrieve, query, automatically mirror, and synchronize data and check data integrity. The NGAS server can be run in three modes: cache server, data-mover (or read-only) server, or regular server. 

The Data Subscription Service of the NGAS enables the synchronization of full or partial data files set to remote Data Subscribers. 
According to data-archiving requirements, the NGAS can be deployed as a Data Provider to send data files and a Data Subscriber in charge of receiving data files. 

\subsection{Transfer Performance Evaluation}
The performance of the NGAS is closely related to network bandwidth and storage system performance. However, network transmission and flow control have more significant impacts on the overall performance. 

\begin{figure*}[htbp]
\centering
\includegraphics[width=0.8\textwidth]{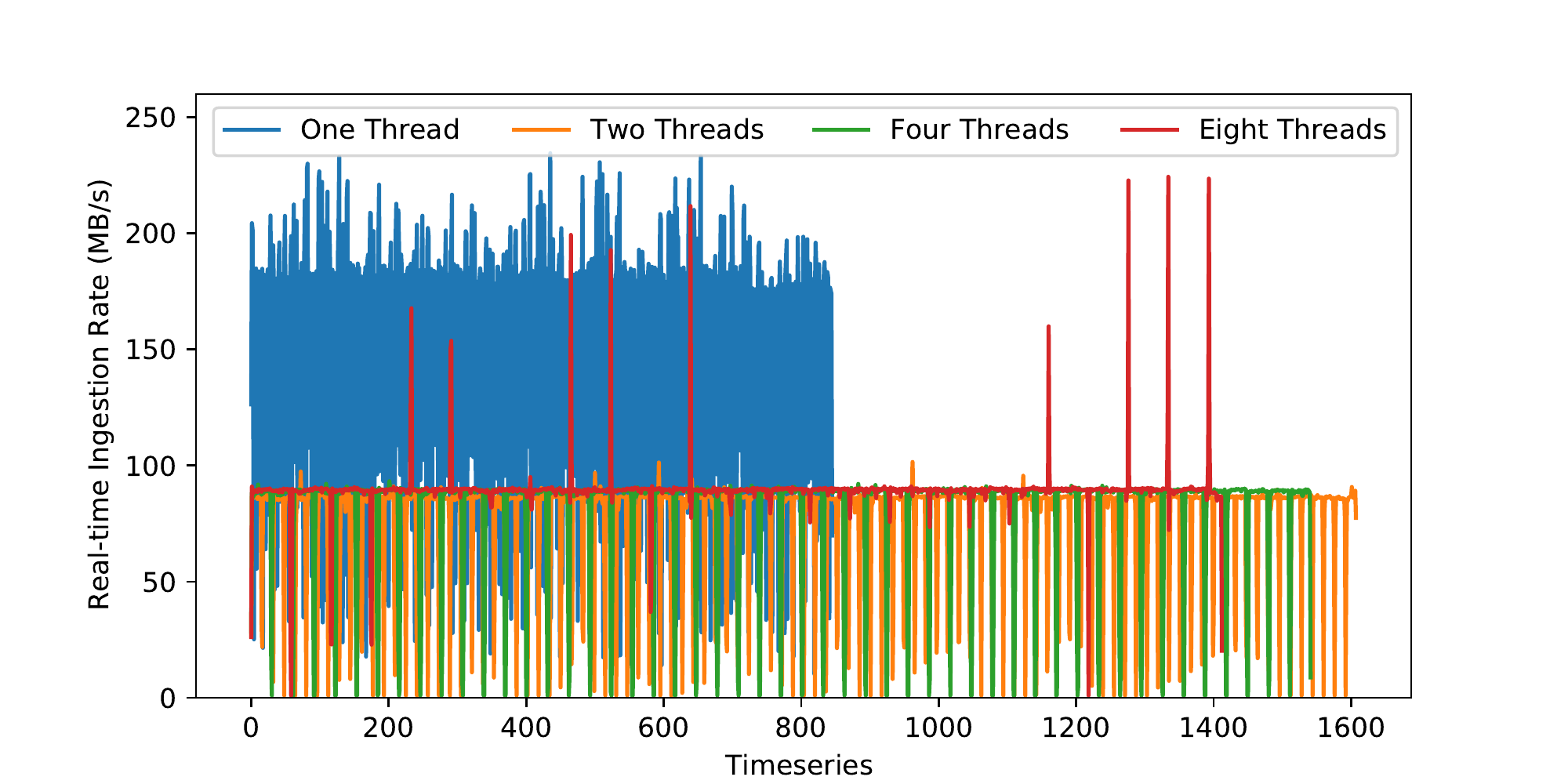}
\caption{Real-time ingestion rates of NGAS under single-process multi-thread mode}\label{FigSingleProcessMultiThreads}
\end{figure*}

We test the performance of the NGAS in the regular server mode according to the construction requirements of the SRC, especially the file transfer performance of the NGAS. Figure~\ref{FigSingleProcessMultiThreads} presents the experimental results of file-transfer performance with different numbers of threads.  
Owing to the HTTP protocol, the NGAS client (Subscriber) could only obtain an average of nearly 160 MB in one thread on a 10 Gb ethernet environment. However, when we attempt to increase the number of threads to boost the transfer performance, the transfer performance decreases instead. 

\begin{figure*}[htbp]
\centering
\includegraphics[width=0.8\textwidth]{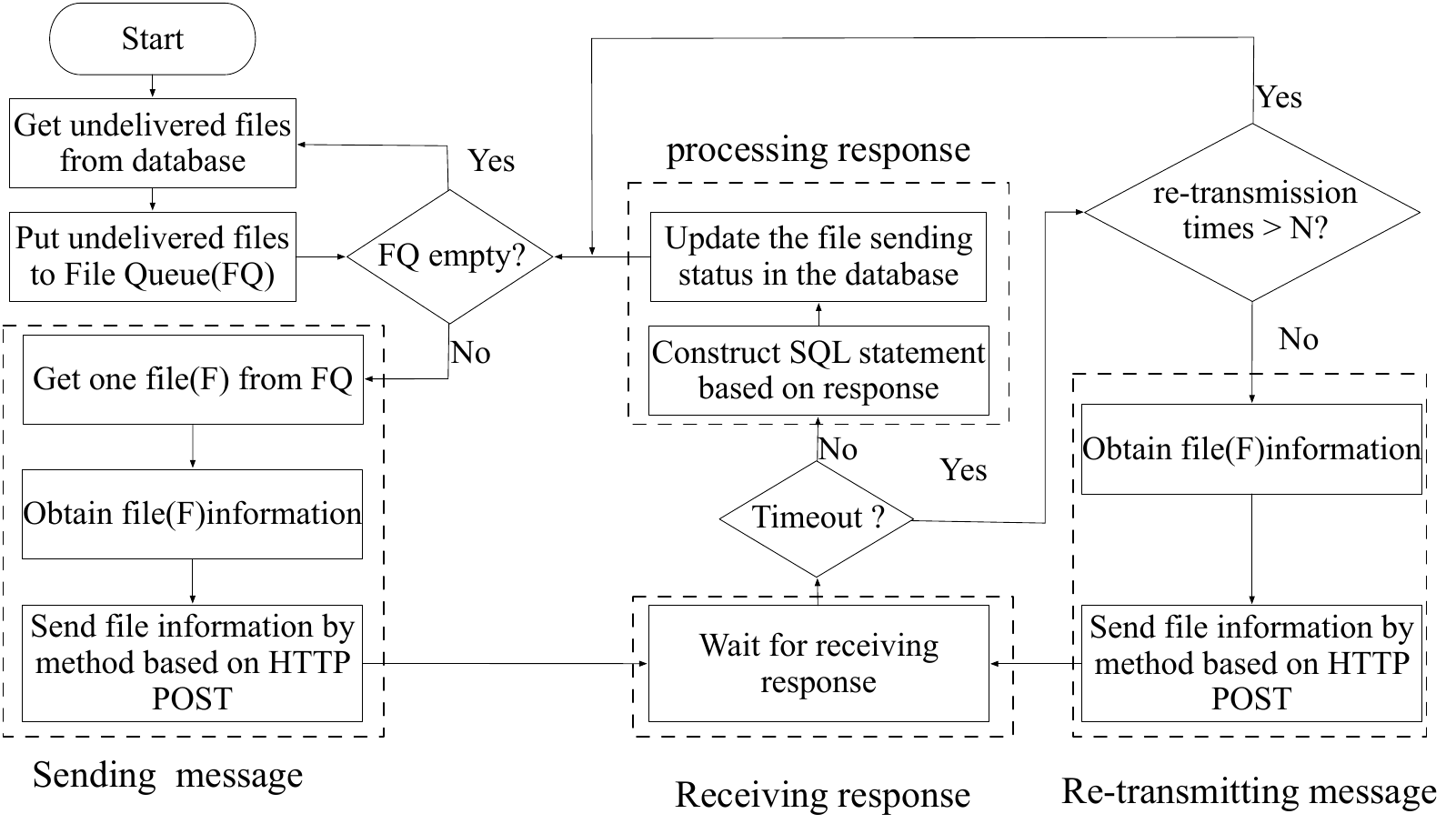}
\caption{Simple diagram of the data delivery thread on the Provider}\label{FigSingleSlidingWindowFlowchart}
\end{figure*}

After the static code analysis of the NGAS, we deduced that the transfer performance could decline owing to two main reasons, as described next. 

1) Flow-control mechanism of NGAS

Figure~\ref{FigSingleSlidingWindowFlowchart} clarifies the delivery procedure of the Provider side.
The delivery procedure mainly includes the send--receive process and re-transmitting sub-procedures.
The send--receive process procedure mainly includes three sub-procedures: sending a message, receiving a response, and processing a response.
The re-transmitting sub-procedure is mainly used to re-transmit the message that has been not received the corresponding response.
The flow-control method used by the data subscription service of the NGAS involves high coupling of the following three sub-procedures in the send--receive process procedure; there is a strict sequence among these sub-procedures. 

\begin{figure*}[htbp]
\centering
\includegraphics[width=1.0\textwidth]{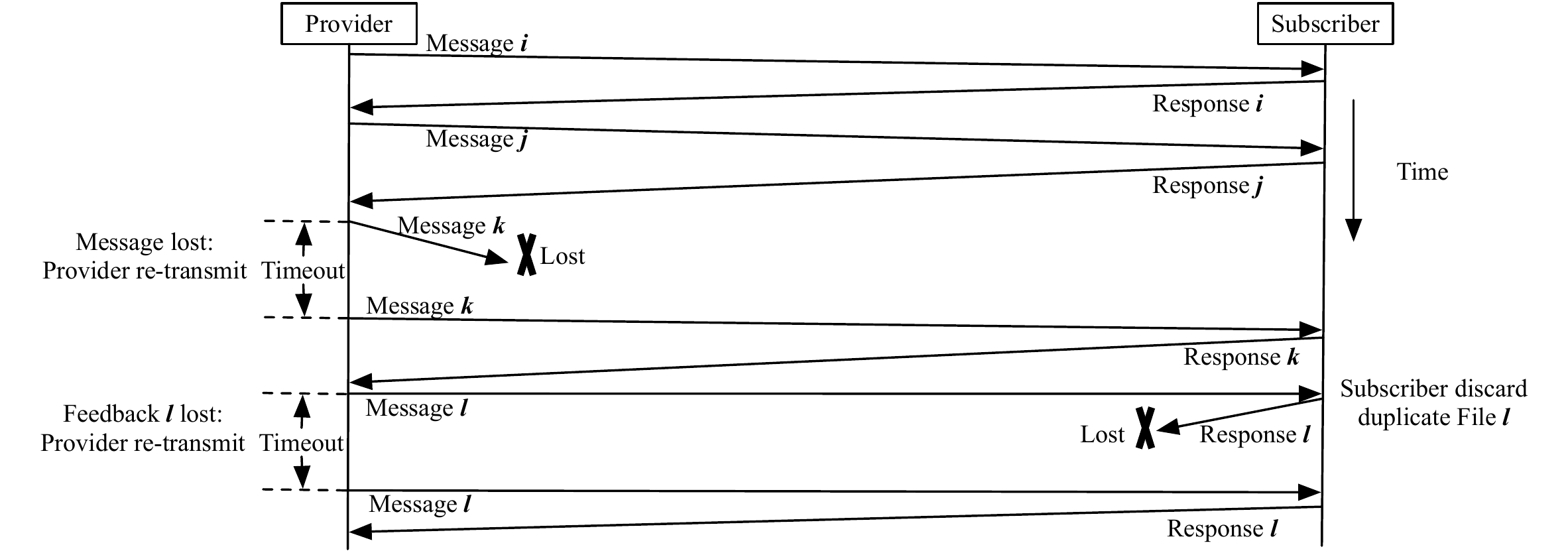}
\caption{Flowchart of NGAS data delivery}\label{FigSingleSlidingWindowConcept}
\end{figure*}

This flow-control method in the data subscription service of the NGAS is similar to the stop-and-wait Automatic Repeat reQuest (ARQ)\citep{bada2017automatic} protocol, with the sliding-window size being 1, shown in Figure~\ref{FigSingleSlidingWindowConcept}. 
In an objective analysis, such a flow-control method is robust. However, it does not allow the Provider to send the following data file to the peer before receiving and processing the peer's response. As a result, the NGAS Provider has to spend considerable time waiting for the response generated by the subscriber till the NGAS Subscriber successfully receives and archives the data file.

2) Limitation of the Python threading module 

Another problem is that the file data delivery process of the Data Subscription Service of the NGAS is implemented using the Python threading module. 
The Python threading module is based on CPython. 
In CPython, owing to the Global Interpreter Lock (GIL), only one thread can execute Python code at once (even though specific performance-oriented libraries might overcome this limitation)\footnote{https://docs.python.org/3.5/library/multiprocessing.html}. 
That means the file data delivery process of the Data Subscription Service of the NGAS will be at the expense of much of the parallelism afforded by multi-processor machines.

\section{Enhanced High-Performance NGAS}\label{sec:e_ngas}

We studied the problem of the relatively low transfer performance of the current NGAS. We first introduce Python's multiprocessing package~\citep{singh2013parallel} to replace the original Python threading module.  We then propose a file-level Unlimited Sliding-Window Technique (USW) based on the Pub/Sub model of ZeroMQ. We also define the message format for high-performance USW communication.  Finally, we redeveloped the NGAS transfer module and implemented an enhanced high-performance archive system (ENGAS).

\subsection{File-level Unlimited Sliding-Window Technique}
\begin{figure*}[htbp]
\centering
\includegraphics[width=1.0\textwidth]{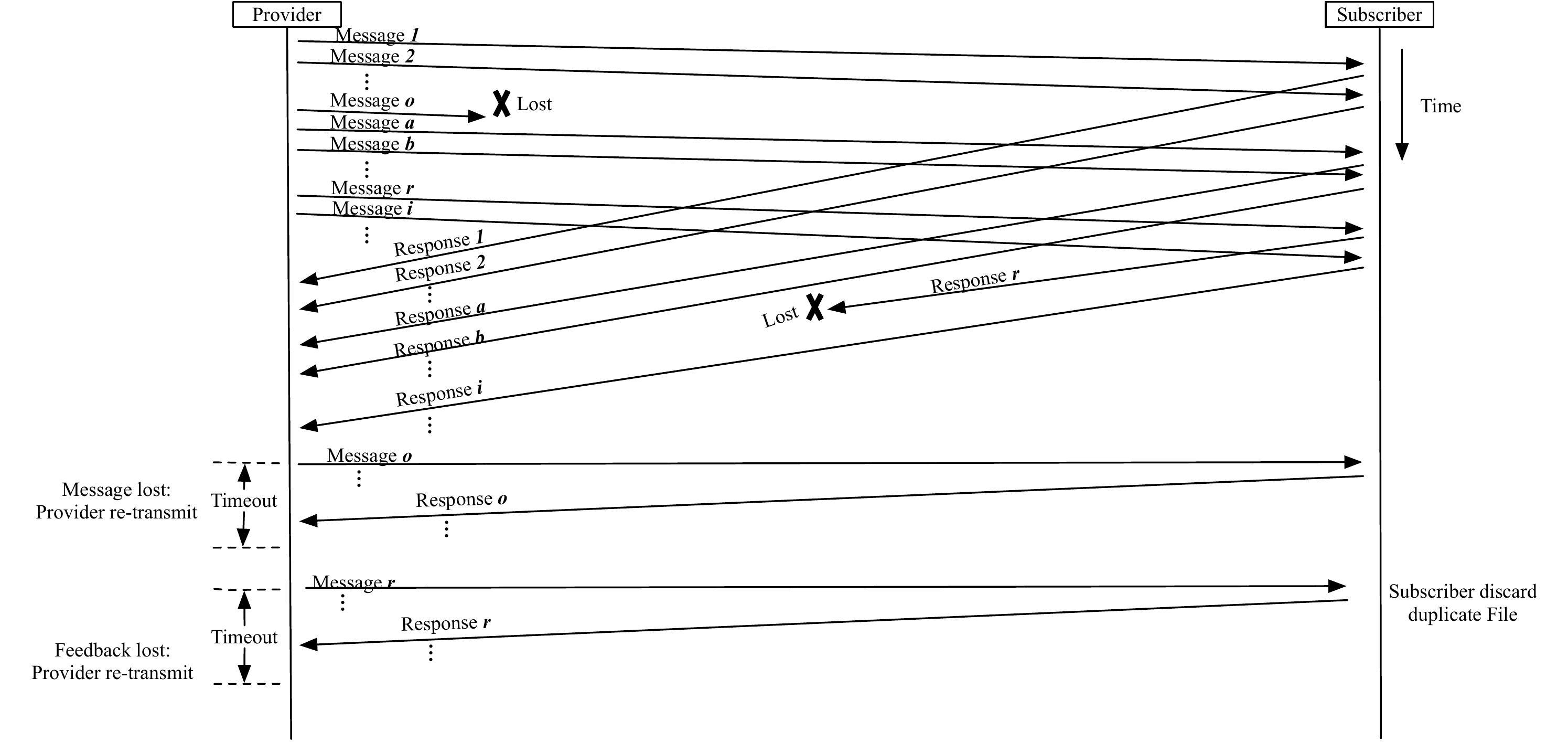}
\caption{Simple flowchart of a file-level USW method for asynchronous data delivery}\label{FigUnlimitedSlidingWindowConcept}
\end{figure*}

Referring to the concept of ARQ that is widely used for error control in data-communication systems\citep{lin1984automatic}, we propose a file-level USW method for asynchronous data delivery upon ZeroMQ communication middleware(refer to  Figure~\ref{FigUnlimitedSlidingWindowConcept}).

\subsubsection{Message Format Definitions in USW}
Two types of messages are designed to satisfy the requirements of the asynchronous communication (Pub/Sub) mode of the ZeroMQ.  The Provider is mainly responsible for sending data files, and the Subscriber is mainly responsible for receiving data files, messages, and responses (refer to Figure~\ref{FigUnlimitedSlidingWindowConcept}).

\begin{figure*}[htbp]
\centering
\includegraphics[width=1.0\textwidth]{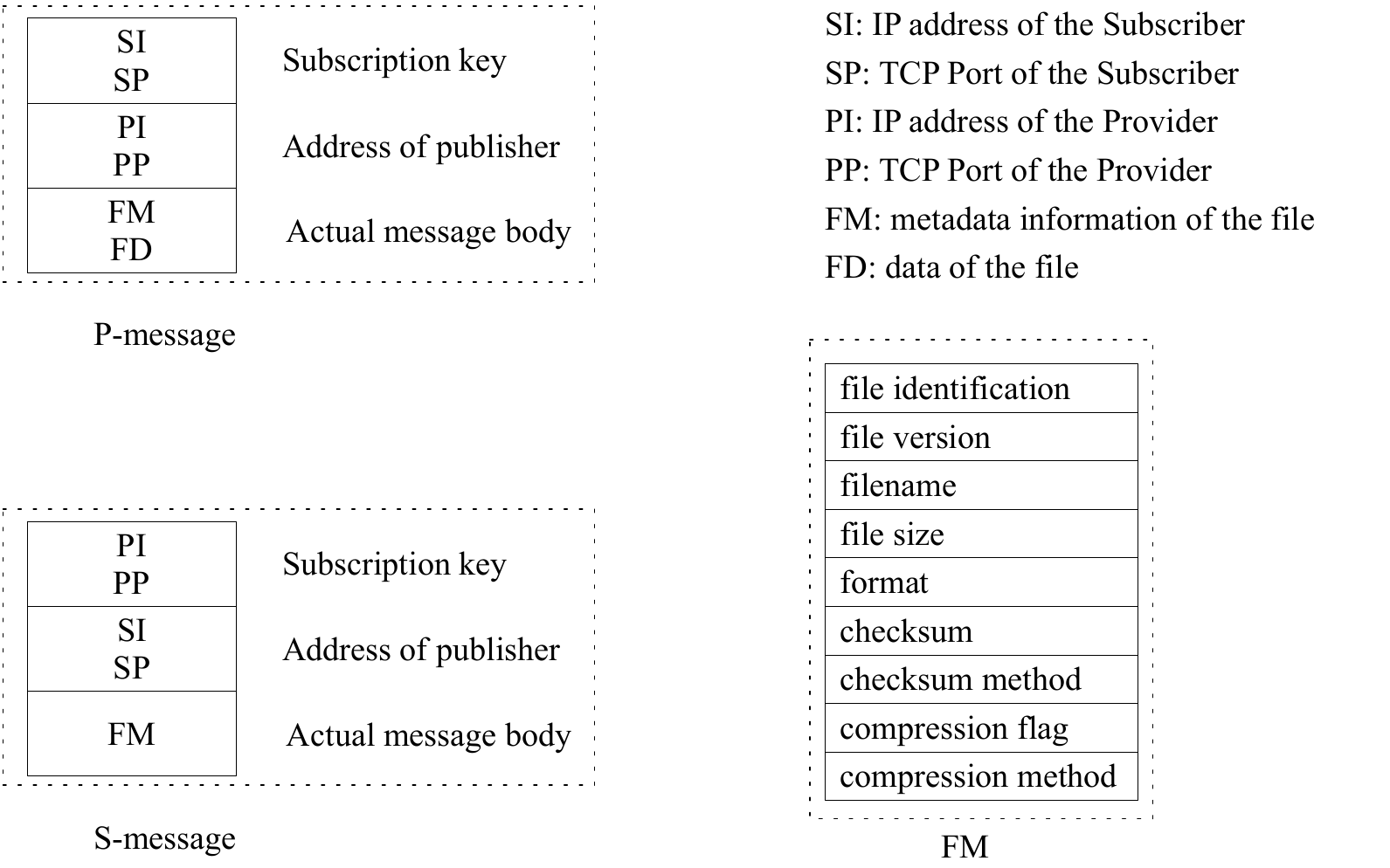}
\caption{Format of P-message and S-message in USW}\label{FigMessageFormat}
\end{figure*}

The P-message (refer to Figure~\ref{FigMessageFormat}) is the formatted message sent by the Provider. The P-message format consists of six parts: IP address of the Subscriber (SI), TCP Port of the Subscriber (SP), IP address of the Provider (PI), TCP Port of the Provider (PP), metadata information of the file (FM), and data of the file (FD). In addition, the FM contains file identification, file version, file size, format, checksum, checksum method, compression flag, and compression method.

The S-message (refer to Figure~\ref{FigMessageFormat}) (response message) is generated by the Subscriber, which consists of five parts: IP address of the Subscriber (SI), TCP Port of the Subscriber (SP), IP address of the Provider (PI), TCP Port of the Provider (PP), and metadata information related to the received file (FM).

\subsubsection{Principle of USW}

The Provider complies with the following: 
1. It records the basic information of the message (SI-SP-PI-PP) for each P-message sent by the Provider.
2. It sends P-messages continuously without waiting for the response message and without considering the storage capacity of the Subscriber.
3. It receives the response message sent by the Subscriber and confirms whether the file has been sent successfully according to the SI-SP-PI-PP information.
4. It re-sends the corresponding file after a timeout error according to the recorded information.

Likewise, the Subscriber complies with the following:
1. It receives the P-message sent by the Provider.
2. It sends an S-message if the file is confirmed as successful.

The USW cannot guarantee that the files sent by the Provider arrive at the Subscriber in that order. However, it can guarantee that the files sent by the Provider eventually arrive at the Subscriber.

\subsection{System Implementation}

Based on the USW and NGAS system, we develop two modules, namely data\_provider and data\_subscriber to replace the origin codes in ngamsServer (refer to Figure~\ref{ngasModuleOrganization}). 

\begin{figure*}[htbp]
\centering
\includegraphics[width=1\textwidth]{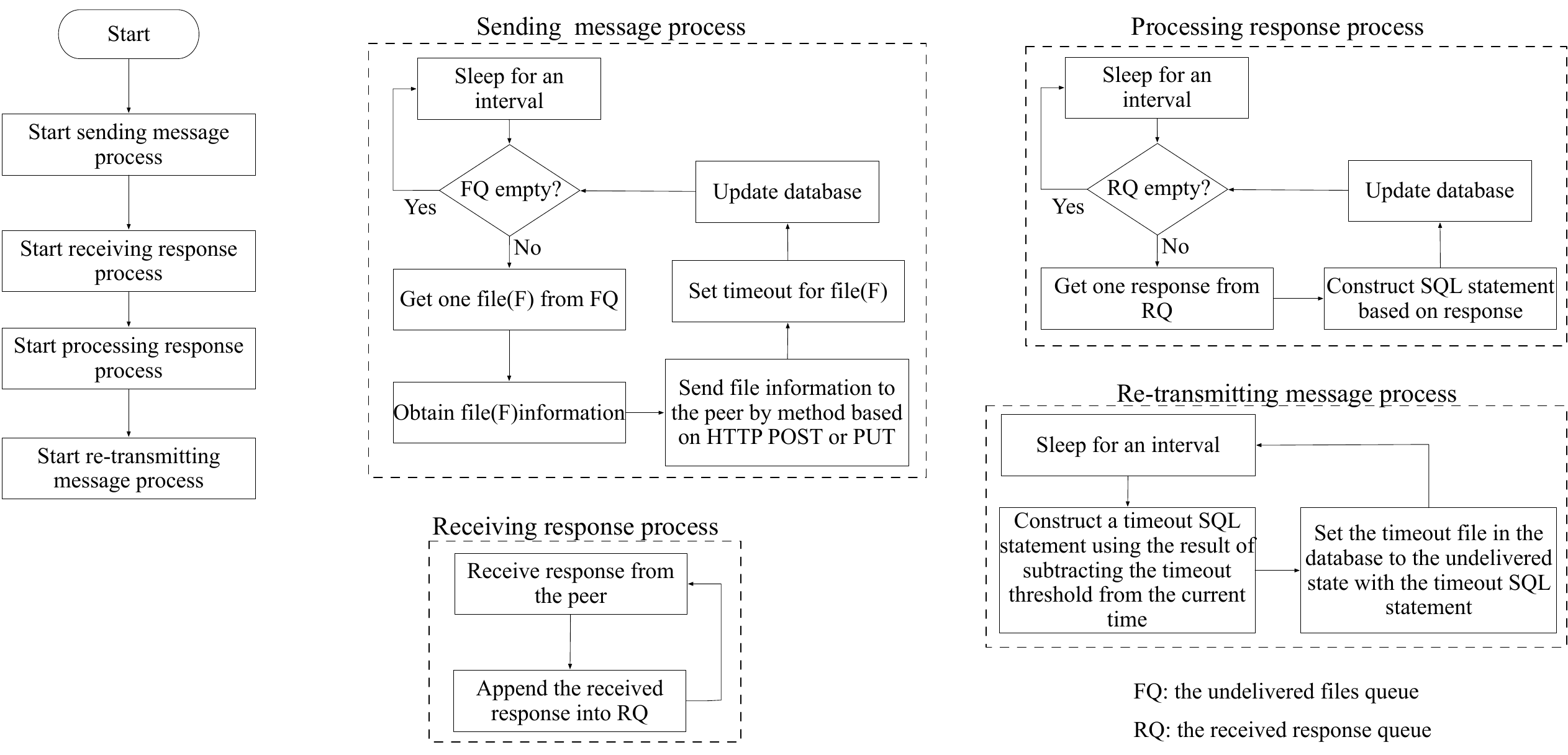}
\caption{Starting sequence of four processes of the ENGAS Provider}\label{FigUnlimitedSlidingWindowFlowchartForProvider}
\end{figure*}

On the one hand, the data\_provider module is implemented for sending messages, receiving response messages, processing response messages, re-transmitting messages, and other functions of the Provider. 
It is deployed as a daemon process. Four main processes will be created and listened the specified ports. (refer to Figure~\ref{FigUnlimitedSlidingWindowFlowchartForProvider}). 

\begin{figure*}[htbp]
\centering
\includegraphics[width=1\textwidth]{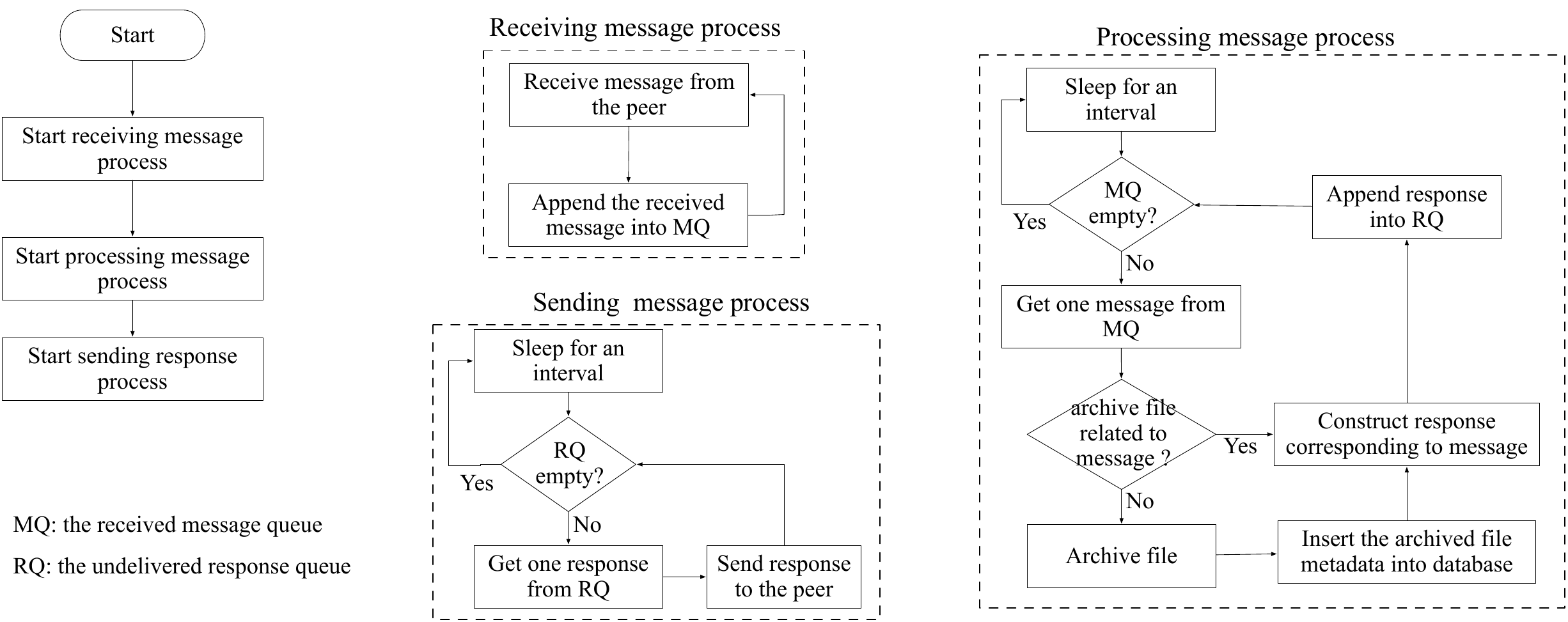}
\caption{Starting sequence of three processes of the ENGAS Subscriber}\label{FigUnlimitedSlidingWindowFlowchartForSubscriber}
\end{figure*}

On the other hand, the data\_subscriber module is implemented for receiving message, processing message, sending response message, and other Subscriber functions as well. It is also deployed as a daemon process in the machines of the Subscribers. Three main processes are created in the system background. (refer to  Figure~\ref{FigUnlimitedSlidingWindowFlowchartForSubscriber}).

After the data\_provider and data\_subscriber modules are initialized, the data\_subscriber module will connect to the pre-defined data\_provider module automatically, and data pairs $<$ sending message process, receiving message process $>$ and $<$ sending response process, receiving response process $>$  will be created based on the ZeroMQ’s Pub/Sub pattern\citep{hintjens2013zeromq}.

\section{Performance Test}\label{sec:experiment}
Two types of tests were designed to compare the performances of ENGAS and NGAS based on MySQL using the InnoDB engine with a default key buffer size of 8 MB, default key cache block size of 1 KB, and default key cache division limit of 300.
The first test compared the performances of ENGAS and NGAS with different file sizes.
The second test compared the performances of ENGAS and NGAS with different concurrencies.

The experimental platform consisted of two Inspur NF5280M5 servers. Each server has two Intel(R) Xeon(R) Silver 4114 CPUs @ 2.20GHz (10 cores for each), 256 GB DDR4-2666ER memory, two 10 Gb Intel Corporation Ethernet adapters, one Samsung 860 SSD drive with 512 GB, and four 2 TB Seagate Enterprise hard drivers. In addition, each server ran on the CentOS 7.4.1708 operating system with MySQL database version 5.6.38, using Python version 3.8.3. During the experiment, one server was deployed as the Provider for sending files, and another server was deployed as a Subscriber in charge of receiving data files.

\begin{table*}[!htb]
\small
\centering
\caption{Experimental data sets\label{tbldatasets}}
\begin{tabular}{llll}
\hline
\hline
 Name & File number & Single file size(Bytes) & Total size (GB)\\
\hline
Dataset1 & 2,000,000 &  66,240 & 123.38\\
Dataset2 & 200,000 &  662,400 & 123.38\\
Dataset3 & 20,000 &  6,624,000 & 123.38\\
Dataset4 & 2,000 &  66,240,000 & 123.38\\
Dataset5 & 200 &  662,400,000 & 123.38\\
Dataset6 & 20 &  6,624,000,000 & 123.38\\
Dataset7 & 333,344 & 397,427 & 123.38\\
\hline
\end{tabular}
\end{table*}

The experimental data sets are based on the simulated data generated by several MWA FITS files downloaded by the Python-based application programming interface (Manta-ray-client\footnote{https://github.com/MWATelescope/manta-ray-client}); 
they are presented in Table~\ref{tbldatasets}.
The total size of each data set was approximately 123.38 GB.

Note that: 1) all experiments were performed in memory; and 2) the average transmission/archiving rate is the total size of each data set divided by the time elapsed. 
The elapsed time is defined as the duration of time between when the first and last files are received.

\subsection{Performance Comparison under Different File Sizes}
To test the impacts of different file sizes on the average transmission performances of the ENGAS and NGAS under the same testing conditions, seven experimental data sets (presented in Table~\ref{tbldatasets}) were used as experimental data. 
The concurrent numbers of ENGAS and NGAS were both set as one in the test.

\begin{figure*}[htbp]
\centering
\includegraphics[width=0.8\textwidth]{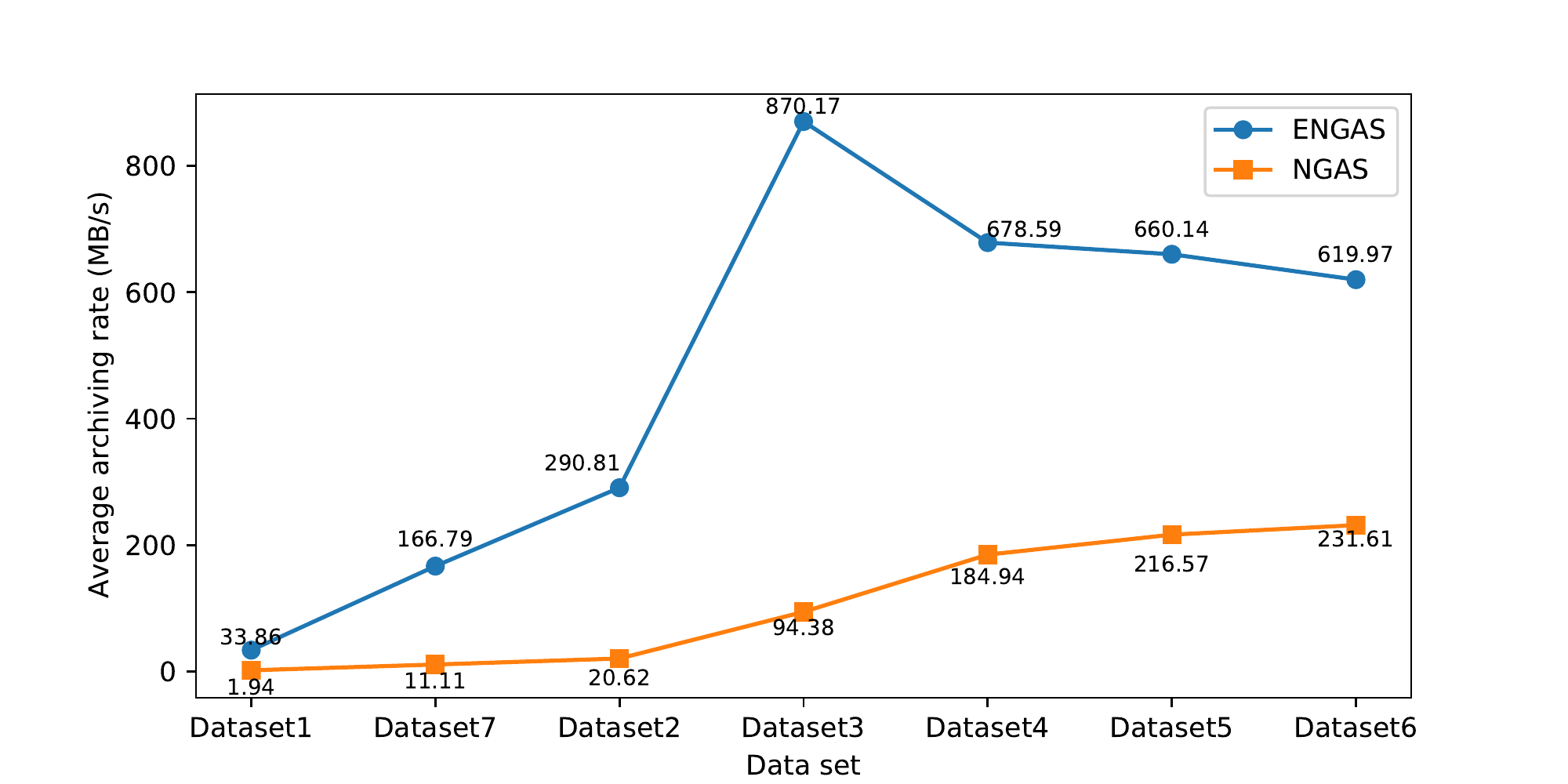}
\caption{Comparison of average transmission rates of ENGAS and NGAS under different file sizes.}\label{FigNewFileSize}
\end{figure*}

The file sizes used in the performance test spanned five orders of magnitude; thus, to demonstrate the relationship between the file size and the average speed more effectively in a relatively small picture, we used the data set of the corresponding file sizes to replace the file sizes (as shown in Figure~\ref{FigNewFileSize}).

As is evident from Figure~\ref{FigNewFileSize}, the average transmission/archiving rate of the ENGAS is faster than that of the NGAS.
When the file size was increased from tens of KB to thousands of MB, the ENGAS went from being nearly 17.45 times faster than the NGAS to being 2.68 times faster.
This phenomenon might have occurred because increasing the file size from tens of KB to thousands of MB decreased the number of files from 2,000,000 to 20, which decreased the overall waiting time for the NGAS.
The proposed file-level USW method for asynchronous data delivery improved the average archiving rate, compared to that for the NGAS. 
In other words, in this situation, the ENGAS could achieve a better data delivery rate than that of NGAS.
Furthermore, for the ENGAS, the file size of Dataset3 was more suitable for remote archives than those of the other six data sets.

\subsection{Performance Variation under Different Concurrencies}
To verify whether different concurrent numbers will affect the average transmission performance, we test the concurrent numbers ranging from 1 to 8 for the ENGAS and NGAS, respectively.
Dataset6 was used as experimental data.

\begin{figure*}[htbp]
\centering
\includegraphics[width=0.8\textwidth]{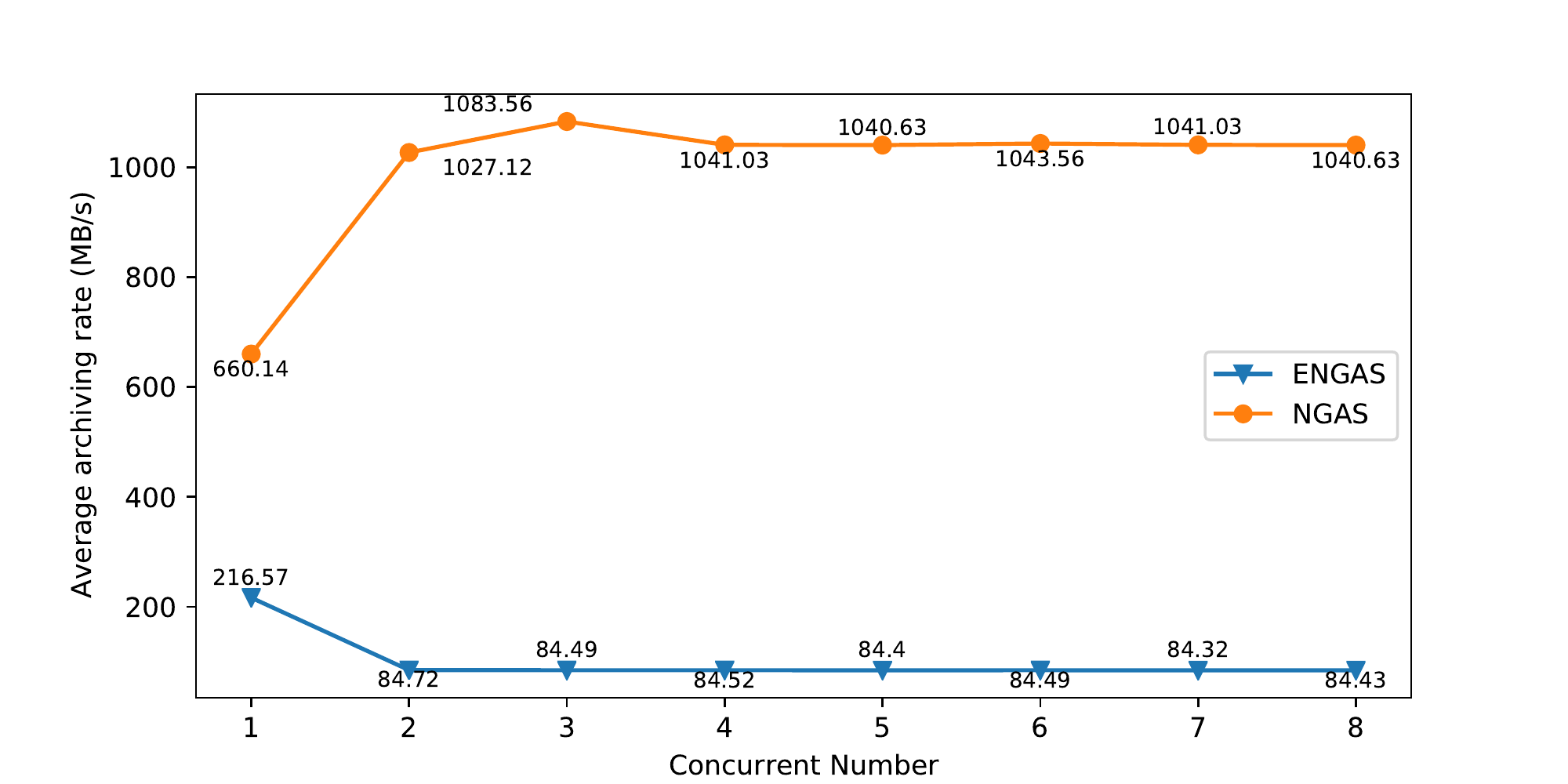}
\caption{Comparison of average ingestion rates of ENGAS and NGAS under different concurrent numbers}\label{FigNewConcurrentNumber}
\end{figure*}

The results are presented in Figure~\ref{FigNewConcurrentNumber}.
Clearly, the average ingestion rate achieved by the ENGAS was significantly faster than that of NGAS.
With an increase in the concurrent number, the average transmission/archiving rate of the ENGAS increased from being almost 3.05 times faster than that of NGAS to approximately 12.33 times faster. 
In addition, the gap between the average archiving rates of the ENGAS and NGAS widened to a fixed value of 12.33.

Furthermore, for the NGAS, with an increase in the number of concurrent threads, the average ingestion rate obtained decreased rapidly from its maximum average ingestion rate (216.57 MB/s) at a concurrent number of 1 to a fixed average ingestion rate (approximately 84.40 MB/s). 
This phenomenon may have occurred because that for a given IO, the multiple threads of the NGAS compete for the resources of a processor, which may cause the average ingestion rate to decrease.

Moreover, as the number of concurrent processes increased, the average archiving rate obtained by the ENGAS initially increased rapidly to its maximum average ingestion rate (1083.56 MB/s) at a concurrent number of 3.
It then decreased gradually, and finally tended to reach a fixed average ingestion rate of approximately 1040.00 MB/s.
Although the maximum average archiving rate and fixed average archiving rate were still lower than the theoretical upper limit (1250 MB/s) at a bandwidth of 10 Gb/s, they were basically equal to the average rate of 1079.00 MB/s tested by iperf3\citep{udayakumar2018bandwidth}.

\section{Discussion}\label{sec:discussion}

Although the ARQ mechanism can guarantee that all the files sent by the Provider eventually arrive at the Subscriber, the order of file transfer is not guaranteed. 
Fortunately, for massive scientific data synchronization between different data centers, the sent data is not required to arrive at the receiving end in strict accordance with the sending order.
Thus, the USW can be used for massive data synchronization in next-generation telescope systems.

Typically, the flow-control mechanism used by the ENGAS is not sufficiently mature. The ENGAS can only achieve file-level re-transmission. Therefore, when the network-link quality is poor and frequent packet loss occurs, the efficiency of the ENGAS becomes very low.

\section{Conclusions}\label{sec:conclusion}
Herein, we proposed a file-level Unlimited Sliding-Window technique (USW) to improve the flow-control performance. 
Experimental results indicate that the USW-based ENGAS delivered acceptable performance in terms of the data transmission rate; the average transmission rate achieved by the ENGAS was faster than that of NGAS (approximately 3.05--12.33 times faster under different experiment conditions; refer to Figure~\ref{FigNewFileSize},\ref{FigNewConcurrentNumber}). The ENGAS is an open-source software that can be freely downloaded and deployed from https://github.com/astronomical-data-processing/ENGAS.

According to the experiment results, the carefully optimized codes of the USW-based ENGAS delivered acceptable archiving/transmission performance.
In this regard, the USW can be considered an effective technique for transmitting/synchronizing massive amounts of data generated by next-generation telescopes, and ENGAS can be considered an effective data transmitting/synchronizing plugin for the NGAS.

\normalem
\begin{acknowledgements}

This work is supported by the National Key Research and Development Program of China (2020SKA0110300), the Joint Research Fund in Astronomy (U1831204, U1931141) under cooperative agreement between the National Natural Science Foundation of China (NSFC) and the Chinese Academy of Sciences (CAS), the National Natural Science Foundation of China (No.11903009). the Funds for International Cooperation and Exchange of the National Natural Science Foundation of China (11961141001), Yunnan Key Research and Development Program(2018IA054). 
The Key Science and Technology Program of Henan Province (No.202102210152, No.212102210611, and No.202102210125),the Research and Cultivation Fund Project of Anyang Normal University (AYNUKPY-2019-24, AYNUKPY-2020-25).

This work is also supported by Astronomical Big Data Joint Research Center, co-founded by National Astronomical Observatories, Chinese Academy of Sciences and Alibaba Cloud. 
The authors would like to thank Chen Wu from International Centre for Radio Astronomy Research, University of Western Australia for providing the valuable suggestions to improve the experiments. The authors gratefully acknowledge the helpful comments and suggestions of the reviewers.

\end{acknowledgements}
  
\bibliographystyle{raa}
\bibliography{bibtex}

\end{document}